\documentstyle[12pt]{article}
\begin{document}
\newcommand{\be}{\begin{equation}}
\newcommand{\ee}{\end{equation}}
\newcommand{\al}{\alpha}
\newcommand{\bt}{\beta}
\newcommand{\lm}{\lambda}
\newcommand{\bea}{\begin{eqnarray}}
\newcommand{\eea}{\end{eqnarray}}
\newcommand{\gm}{\gamma}
\newcommand{\Gm}{\Gamma}
\newcommand{\dl}{\delta}
\newcommand{\Dl}{\Delta}
\newcommand{\ep}{\epsilon}
\newcommand{\kp}{\kappa}
\newcommand{\Lm}{\Lambda}
\newcommand{\om}{\omega}
\newcommand{\pa}{\partial}
\newcommand{\dd}{\mbox{d}}
\newcommand{\MS}{\mbox{MS}}
\newcommand{\nn}{\nonumber}
\newcommand{\uk}{\underline{k}}

\title{
 \vspace*{-30mm}
 \begin{flushright}
{\normalsize  TTP/99--54\\[1mm] 
DESY/00--008    \\[-0mm]
hep-ph/9912503    \\[-3mm]
December 1999}    \\[30mm]
 \end{flushright}
Summing up Subleading Sudakov Logarithms} 
\author{
  J.H.~K\"uhn\\[-0mm] 
  {\small {\em Institut f{\"u}r Theoretische Teilchenphysik,
  Universit{\"a}t Karlsruhe}}\\[-2mm]
  {\small {\em 76128 Karlsruhe, Germany}}\\[2mm]
  A.A.~Penin\thanks{On leave from  Institute for Nuclear 
  Research of Russian Academy   of Sciences,
  117312 Moscow, Russia}\\[-0mm]
  {\small {\em II. Institut f{\"u}r Theoretische Physik,
  Universit{\"a}t Hamburg}}\\[-2mm]
  {\small {\em  22761 Hamburg, Germany}}\\[2mm]
  V.A.~Smirnov \\[-0mm] 
  {\small {\em  Nuclear Physics
  Institute of Moscow State University}}\\[-2mm]
  {\small {\em  119899 Moscow, Russia}}
}

\date{}
\maketitle

\begin{abstract}
We apply the strategy of regions within dimensional
regularization to find functions involved in evolution
equations which govern the asymptotic dynamics of the 
Abelian  form factor and four-fermion
amplitude in the $SU(N)$ gauge theory  in the Sudakov 
limit up to the next-to-leading logarithmic approximation.
The results are used for  the analysis of the 
dominant electroweak corrections to the fermion-antifermion pair production in the $e^+e^-$ annihilation
at high energy.  
\\[2mm]
PACS numbers:  12.38.Bx, 12.38.Cy, 12.15.Lk 
\end{abstract}

\thispagestyle{empty}

\newpage

\section{Introduction}

The asymptotic behavior of various amplitudes
in the Sudakov limit 
has been investigated in QED and QCD, with summation of
the leading double \cite{Sud,Jac,CorTik,FreTay}
and subleading   \cite{Smi,Col,Sen1,Sen2,Kor1,Kor2,Mag}
logarithms. Evolution equations that govern
the dynamics of the amplitudes in the Sudakov limit 
have been obtained in refs.~\cite{Col,Sen1,Sen2,Kor2}. 
In the present paper, we apply this approach to the
next-to-leading analysis of the  Abelian form factor 
and the four fermion amplitude in the $SU(N)$ gauge theory.
We evaluate functions that enter the evolution equations
in the  next-to-leading logarithmic approximation by using, as an
input, asymptotic expansions of one-loop diagrams.
Solving these equations we sum  up the leading
and subleading Sudakov logarithms.
The  expansions of one-loop diagrams are obtained by use of the so-called
generalized strategy of regions \cite{BS} (see also \cite{SR})
which enables us to systematically identify the nature
of various contributions and the origin of logarithms.
This strategy is based on expanding 
integrands of Feynman integrals in typical regions and 
extending the integration domains to the whole space of
the loop momenta so that a crucial difference with respect to
the standard approach \cite{Col,Sen1,Sen2,Kor1,Kor2} is the absence
of cut-offs that specify the regions in individual terms of 
the expansions. 
This approach is applied within
dimensional regularization \cite{dimreg} when all 
integrals without scale (not only massless vacuum integrals)
are put to zero. 

We apply our  results for the subleading Sudakov logarithms 
to the analysis of the  dominant electroweak  corrections  to the 
process of the fermion-antifermion pair production
in the $e^+e^-$ annihilation.  
In the standard model of weak interactions 
the $W$ and $Z$ bosons get their masses  
via the Higgs mechanism and the Sudakov logarithms naturally
appear in the virtual electroweak corrections \cite{KDB}.
They grow rapidly with energy and become
dominant  in the TeV region available at the 
LHC or the Next Linear Collider. 
The analysis of the Sudakov corrections 
is thus of high importance for the next generation of accelerators.
The leading and subleading electroweak Sudakov logarithms
were discussed in \cite{CiaCom,Bec} in one loop approximation. 
The effect of
higher order  leading logarithms was estimated in 
\cite{KuhPen} by computing, in a physical gauge, 
a contribution related to the  multiple virtual $Z$ and $W$ bosons 
exchanges. 
A complete analysis of the  leading logarithms in exclusive and
inclusive electroweak processes was done in  \cite{Fad} on the 
basis of the infrared evolution equation approach.
The subleading Sudakov logarithms, however, turn out to be significant
and should be taken into account to get a reliable estimate of high 
order corrections.

The paper is organized as follows. 
In the next section, the Abelian form factor is analyzed 
in two typical Sudakov type regimes. 
The analysis is then extended in Section~3 to the four-fermion amplitude.
The electroweak Sudakov corrections  are considered in Section~4.
We present our conclusions in the last section.

\section{The  Abelian form factor in the Sudakov limit}

The (vector) form factor  which determines the  amplitude of the
fermion scattering in the external Abelian field  in the Born 
approximation can be written as follows
\be
F_B=\bar\psi(p_2)\gm_\mu\psi(p_1)\; ,
\ee
where the four-vector index on the left hand side is suppressed,
$p_1$ is incoming and $p_2$ is outgoing momentum.

There are two ``standard'' regimes of the Sudakov limit
$s=(p_1-p_2)^2 \to-\infty$ \cite{Sud,Jac}:

({\em i}) On-shell massless fermions, $p_1^2=p_2^2=0$, and
gauge bosons with a small non-zero mass $M^2\ll -s$.
Let us choose, for convenience,
$p_{1,2} = (Q/2,0,0,\mp Q/2)$ so that  $2 p_1 p_2 = Q^2=-s$.

({\em ii}) The massless gauge bosons and off-shell massless fermions
$p_1^2=p_2^2=-M^2$, $M^2\ll -s$.
We choose
$p_1 = \tilde{p}_{1,2} -(M^2/Q^2) \tilde{p}_{2,1}$,
where $\tilde{p}_{1,2}$ are defined as $p_{1,2}$ in the regime~({\em i}).

The asymptotics of the form factor in  the Sudakov limit
can be found by solving the corresponding evolution equation \cite{Col}. For
the non-Abelian gauge theory, this  equation was first derived
in \cite{Sen1} by factorizing collinear logarithms in the 
axial gauge.    In the first regime, it reads
\be
{\partial\over\partial\ln{Q^2}}F=
\left[\int_{M^2}^{Q^2}{\dd x\over x}\gm(\al(x))+\zeta(\al(Q^2))
+\xi(\al(M^2)) \right] F \; .
\label{eei}
\ee
Its solution is
\be
F=F_0(\al(M^2))\exp \left\{\int_{M^2}^{Q^2}{\dd x\over x}
\left[\int_{M^2}^{x}{\dd x'\over x'}\gm(\al(x'))+\zeta(\al(x))
+\xi(\al(M^2))\right]\right\} \; .
\ee
A generalization of eq.~(\ref{eei}) to the regime ({\em ii}) was found in
\cite{Kor2}:
\be
{\partial\over\partial\ln{Q^2}}F=
\left[\int_{M^4/Q^2}^{Q^2}{\dd x\over x}\gm(\al(x))+
\zeta(\al(Q^2))+ \zeta'(\al(M^4/Q^2))+\xi(\al(M^2))\right] F\; .
\ee
Its solution is
\bea
F&=&F_0(\al(M^2))\exp\left\{\int_{M^2}^{Q^2}{\dd x\over x}
\left[\int_{M^2}^{x}{\dd x'\over x'}\gm(\al(x'))+\zeta(\al(x))
+\xi(\al(M^2))\right]\right.
\nn \\ &&
\left.
+ \int^{M^2}_{M^4/Q^2} {\dd x\over x}
\left[\int^{M^2}_{x}{\dd x'\over x'}\gm(\al(x'))+\zeta'(\al(x)) \right]
\right\} \, . 
\eea
The functions  $F_0$ and   $\xi$ are, generally, different in the
two regimes.
We are interested in the next-to-leading  logarithms. Therefore
we should keep renormalization group corrections to the leading
logarithmic approximation
as well as single infrared and renormalization group
logarithms. In this approximation, the  form factor ({\em i}) takes the form
\be
F=F_0(\al)\exp\left[\int_{M^2}^{Q^2}{\dd x\over x}
\int_{M^2}^{x}{\dd x'\over x'}\gm(\al(x'))+(\zeta(\al)
+\xi(\al))\ln{(Q^2/M^2)}\right]
\ee
and, in the regime ({\em ii}), we have
\[
F=F_0(\al)\exp\left\{\int_{M^2}^{Q^2}{\dd x\over x}
\int_{M^2}^{x}{\dd x'\over x'}\gm(\al(x'))
+\int^{M^2}_{M^4/Q^2} {\dd x\over x}
\int^{M^2}_{x}{\dd x'\over x'}\gm(\al(x'))\right.
\]
\be
+(\zeta(\al)
+\zeta'(\al)+\xi(\al))\ln{(Q^2/M^2)}\bigg\} \; .
\ee
All the functions in the exponent have to be computed in one loop,
and the one loop running of $\al$ in the argument of the
$\gm$ function should be taken into account.   
Note that in the next-to-leading order we cannot separate
the functions $\zeta$  and $\xi$ but we will see that $\zeta'$
vanishes in one loop.

In the covariant gauge, the self energy insertions
to the external fermion lines do not give $Q$-dependent
contributions.   
The one loop calculation of the vertex correction gives
\be
F={\al\over 2\pi}C_F\left(-V_0+2V_1+2(1-2\ep)V_2-V_2'\right)F_B\, ,
\ee
where $C_F=(N^2-1)/(2N)$ is the  quadratic Casimir operator of
the fundamental representation of $SU(N)$ group  
and the functions involved in  are given by  
\bea
\int \frac{\dd^d k }{(k^2-2p_1k) (k^2-2p_2k) (k^2-M^2)} &=& 
i\pi^{d/2}e^{-\gm_{\rm E}\ep} s^{-1} V_0  \; ,
\nn \\
\int \frac{\dd^d k \; k_\mu}{(k^2-2p_1k) (k^2-2p_2k) (k^2-M^2)} &=&
i\pi^{d/2}e^{-\gm_{\rm E}\ep} s^{-1} (p_1+p_2)_\mu V_1 \; ,
\nn \\
\int \frac{\dd^d k \; k_\mu k_\nu}{(k^2-2p_1k) (k^2-2p_2k) (k^2-M^2)}
&=&
i\pi^{d/2}e^{-\gm_{\rm E}\ep} 
\left[g_{\mu\nu}V_2+{{p_1}_\mu{p_2}_\nu
+ (\mu\leftrightarrow\nu)\over s}V_2' \right] \; .
\label{ints}
\eea
in  the regime ({\em i}). 
The corresponding  relations for  
the regime ({\em ii}) are obtained from eq.~(\ref{ints}) by the 
following substitution
\be
(k^2-2p_1k) (k^2-2p_2k) (k^2-M^2) \to (k^2-2p_1k-M^2) (k^2-2p_2k-M^2) k^2 \; ,
\ee 
with another set of the functions $V$ involved.
We omit $i 0$ in $k^2-2p_1k+i 0$, etc. for brevity.
Here $\gm_{\rm E}$ is the Euler constant.
We work in dimensional regularization \cite{dimreg} with $d=4-2\ep$.
We also usually omit the factor $(\mu^2)^{\ep}$ per loop
and write it down only in the argument of the renormalization group  logarithm.
Note that only the tensor structures giving unsuppressed  
contributions to the  form factor are kept in the representation 
of the third integral.

To expand these integrals in the limit $Q^2\gg  M^2$ we apply a generalized
strategy of regions formulated in \cite{BS} and
discussed using characteristic two-loop examples in~\cite{SR}:
\begin{itemize}
\item Consider various regions of the loop momenta and expand, in
every region, the integrand in Taylor series with respect to the
parameters that are there considered small;
\item Integrate the  expanded integrand
over the whole integration domain of the loop momenta;
\item Put to zero any scaleless integral.
\end{itemize}

The following regions happen to be typical in the Sudakov 
limit \cite{Ste}:
\bea
\label{h}
\mbox{{\em hard} (h):} && k\sim Q\, ,
\nn \\
\label{1c}
\mbox{{\em 1-collinear} (1c):} && k_+\sim Q,\,\,k_-\sim M^2/Q\, ,
\,\, \uk \sim M\,,
\nn \\
\label{2c}
\mbox{{\em 2-collinear} (2c):} && k_-\sim Q,\,\,k_+\sim M^2/Q\, ,
\,\,\uk \sim M \, ,
\nn \\
\label{soft}
\mbox{{\em soft} (s):} && k\sim M \, ,\nn \\ 
\label{us}
\mbox{{\em ultrasoft} (us):} && k\sim M^2/Q\, .
\nn 
\eea
Here $k_{\pm} =k_0\pm k_3, \; \uk=(k_1,k_2)$. We mean by $k\sim Q$, etc.
that any component of $k_{\mu}$ is of order $Q$.

Keeping the leading power in the expansion in the limit
$Q^2/M^2\to \infty$ we have in the regime 
({\em i})\footnote{In fact we do not need, in
the LO and NLO, finite parts present in these and similar 
results. They would be, however, needed for the NNLO calculations.}
\bea
V_0^h&=&{1\over \ep^2}-{1\over \ep}\ln{(-s)}
+{1\over 2}\ln^2{(-s)}-{\pi^2\over 12} \; ,\nn \\
V_0^c&=&-{1\over \ep^2}+{1\over \ep}\ln{(-s)}
-\ln{(M^2)}\ln{(-s)}+{1\over 2}\ln^2{(M^2)}+{5\pi^2\over 12} \; , \\
V_0&=&{1\over 2}\ln^2\left({-s\over M^2}\right)+{\pi^2\over 3} \; ;
\nn 
\eea
\bea
V_1^h&=&-{1\over \ep}+\ln{(-s)}-2 \; ,\nn \\
V_1^c&=&{1\over \ep}-\ln{(M^2)}+1 \; , \label{v1i}\\
V_1&=&\ln\left({-s\over M^2}\right)-1 \; ;
\nn 
\eea
\bea
V_2=V_2^h&=&{1\over 4}\left({1\over \ep}-
\ln\left({-s\over\mu^2}\right)+3\right) \; , \nn \\
V_2'={V_2'}^h&=&{1\over 2} \; .
\label{v2}
\eea
We denote by the index $c$ the sum of the 1c and 2c contributions.  
The  pole in eq.~(\ref{v2}) is of  ultraviolet nature. It is not canceled
by the collinear or ultrasoft contributions. 
Note that the corresponding ultraviolet renormalization group 
logarithm $\ln\left(-s/\mu^2\right)$ contributes only to 
the function $\zeta$.

The hard part in the regime ({\em ii})  is the same. The new ingredients
read
\bea
V_0^c~&=&-{2\over \ep^2}+{2\over \ep}\ln{(M^2)}-\ln^2{(M^2)}+{\pi^2\over 6}
\; ,\nn \\
V_0^{us}&=&{1\over\ep^2}+{1\over \ep}\left(\ln{(-s)}-2\ln{(M^2)}\right)
+{1\over 2}\ln^2{(-s)}
\nn \\ &&
-2\ln{(M^2)}\ln{(-s)}+2 \ln (M^2) +{\pi^2\over 4} \; , \label{v0ii} \\
V_0~&=&   \ln^2\left({-s\over M^2}\right)+{\pi^2\over 3} \; ;
\nn 
\eea
\bea
V_1^c&=&{1\over \ep}+2-\ln{(M^2)} \; ,\nn \\
V_1&=&\ln\left({-s\over M^2}\right) \; .
\label{v1ii}
\eea
Soft regions generate
only zero contributions (at least in the leading power).

 From the one-loop result  we find  
\be
\gm(\al)=-C_F{\al\over 2\pi} \;.
\ee
Moreover it is clear from the above expressions  that
in the regime ({\em i}) the total double logarithms of $Q$
come from the hard region  while in the regime ({\em ii})
one half of the double logarithmic contribution
comes from the ultrasoft region. This explicitly determines
the scale of the coupling constant in the second order
logarithmic derivative of the form factor in $Q$. It is $Q$
in the regime ({\em i}) and  $M$ in the regime ({\em ii}).
Furthermore, all  the $Q$ dependent terms in the ultrasoft contribution
of eq.~(\ref{v0ii}) are related to the $\gm$ term
and therefore  $\zeta'(\al)=0$. 
At the same time we cannot distinguish, in the one loop approximation, 
the contribution to the functions $\zeta$  and $\xi$ 
coming from the collinear region  
because this region includes both $Q$ and $M$ scales.
For the sum of these functions we find
\be
\zeta(\al)+\xi(\al)=3C_F{\al\over 4\pi} \; .
\ee
To complete the $Q$-independent part of the one loop corrections  to the
form factor one has  to include the fermion wave function
renormalization determined by the self energy insertions
to the external  lines. In the regime ({\em i}) this
brings the factor
\be
1+C_F{\al\over 4\pi}
\left(-{1\over \ep}+\ln{\left({M^2\over\mu^2}\right)}+{1\over 2}\right)
\label{wfi}
\ee
and in  the regime ({\em ii}) this gives
\be
1+C_F{\al\over 4\pi}
\left(-{1\over \ep}+\ln{\left({M^2\over\mu^2}\right)}-1\right) \;.
\label{wfii}
\ee
The ultraviolet poles of eqs.~(\ref{wfi},~\ref{wfii}) cancel the  
ultraviolet pole in the first line of eq.~(\ref{v2}) due to Ward identity
and nonrenormalization property of the conserved vector current.
Finally, in the NLO logarithmic approximation, we find 
the form factor ({\em i}) to be
\bea
F &=& F_B\left(1-C_F{\al\over 2\pi}\left({7\over 2}+{2\pi^2\over 3}\right)\right)
\exp \left\{{C_F\over 2\pi}\left[-\int_{M^2}^{Q^2}{\dd x\over x}
\int_{M^2}^{x}{\dd x'\over x'}\al(x')\right.\right.
\nn \\ && 
+3\al\ln{(Q^2/M^2)}\bigg]\bigg\}
\label{fi}
\eea
and  the form factor ({\em ii}) to be 
\bea
F&=&F_B\left(1-C_F{\al\over 2\pi}\left(1+{2\pi^2\over 3}\right)\right)
\exp\left\{{C_F\over 2\pi}\left[-\int_{M^2}^{Q^2}{\dd x\over x}
\int_{M^2}^{x}{\dd x'\over x'}\al(x')
\right. \right.
\nn \\ && 
-\int^{M^2}_{M^4/Q^2} {\dd x\over x}
\int^{M^2}_{x}{\dd x'\over x'}\al(x')
+3\al\ln{(Q^2/M^2)}\bigg]\bigg\} \; .
\label{fii}
\eea
The single  logarithmic term in the exponents of 
eqs.~(\ref{fi},~\ref{fii}) has the following  decomposition 
\be
3\ln{(Q^2/M^2)}=4\ln{(Q^2/M^2)}_{(IR)}-\ln{(Q^2/M^2)}_{(RG)}\; ,
\label{decomp}
\ee
where we explicitly separate the infrared and renormalization 
group\footnote{Although
this logarithm originates from the  integration over the virtual momentum 
region between
$M$ and $Q$ scales and does not depend on $\mu$ we 
call it ``renormalization group'' one because it is directly
related and can be read off the renormalization group properties of 
the Abelian vertex and the fermion wave function.}  
logarithms which are related to $V_1$ and $V_2$ integrals
correspondingly.
These logarithms  are of essentially 
different nature (see a discussion below).
Eqs.~(\ref{fi},~\ref{fii}) are in agreement with the result of 
refs.~\cite{Kor1,Kor2}.

It is useful to distinguish the soft and collinear 
poles in $\ep$ (resulting in logarithms)
in the hard contribution. The collinear logarithms in a physical
(Coulomb or axial) gauge originate only from the self energy
insertions into the external fermion  lines 
\cite{FreTay,Col,Sen1,Sen2,FreMeu,Fre,Ama}  and
therefore are universal {\it i.e.} independent of specific
processes.  In contrast to the soft divergences (i.e. infrared
divergences that are local in momentum space),
the divergences of this type arise from the
integration over angle variables. Consider, for example,
the integral $V_1$.
The power counting tells us that
there is no soft divergence in this integral (we have $k^5$
in the numerator and only $k^4$ in the denominator).
However, for non-zero $k$ with $k^2\sim 0$ which is
collinear to $p_1$ or $p_2$ (when $p_1k\sim 0$ or $p_2k\sim 0$),
the integrand blows up. This follows from the fact that
the factor $1/((k^2-2pk)k^2)$, with any $p^2=0$, generates collinear
divergences. When integrating this factor in $k_0$, take
residues in the upper half plane. For example, taking a residue
at $k_0=-|\vec{k}|$ leads to an integral with
$1/(pk)=1/(p^0 |\vec{k}| (1-\cos\theta))$ where $\theta$ is the angle
between the spatial components. Thus, for  small $\theta$,
we have a divergent integration over angles
$d \cos\theta/(1-\cos\theta)\sim d\theta/\theta$.
The second residue generates a similar divergent behaviour ---
this can be seen by the change $k\to p-k$.

Within the method of regions  the total divergence
of the collinear  region in general cancels both the soft
and  collinear poles of the hard part. Hence
it is not straightforward to separate the collinear logarithms. 
Let us note that the collinear divergences (but not the
soft ones) can be regularized by introducing a finite fermion 
mass. The reason is that the light-like vector $k$ cannot
be collinear to the space-like vector $p_i$ in this case.
This enables us to distinguish the soft and collinear poles.
Introducing a small fermion mass $m$ but
keeping zero gauge boson mass we get the  collinear poles in the
hard part canceled by the poles of the contribution
from the collinear region leaving the
logarithms while the soft poles in the hard part are not canceled.
Thus one can determine the origin of the poles in the hard part
and  therefore the origin of the logarithms.
For example,  in the hard part of the integral $V_1$
the pole  is of the   collinear
origin since it is canceled in the regime $m\ne 0$, $M=0$
by
\be
V_1^c={1\over \ep}-\ln{(m^2)}+2 \; .
\label{v1c}
\ee
Thus the single  infrared logarithm 
in eq.~(\ref{decomp})
is of collinear origin and therefore is universal.
We should emphasize  that this is not true for the
renormalization group  logarithm of this equation
which depends on a specific
amplitude and a model. 
For example it is different for the scalar form factor
or for the vector form factor
in a model with an additional Yukawa interaction of the fermions
with the scalar bosons.

A less trivial example is the integral $V_0$.
In the hard part of $V_0$ in this regime,
the collinear part of the double pole is canceled by
\be
V_0^c=-{1\over \ep^2}+{1\over \ep}\ln{(m^2)}-{1\over 2}\ln^2{(m^2)}
-{\pi^2\over 12}
\ee
and transforms into the logarithm of $s/m^2$ but the soft
single pole is left
\be
V_0=-{1\over \ep}\ln\left({-s\over m^2}\right)
+\ln\left({-s\over m^2}\right)\ln{(-s)}
-{1\over 2}\ln^2\left({-s\over m^2}\right)-{\pi^2\over 6} \; .
\ee

\section{The four fermion amplitude}
We study  the limit of the fixed-angle scattering when 
all the  invariant energy and momentum transfers
of the process are much larger than the 
typical mass scale of internal particles $|s|\sim |t| \sim |u| \gg M^2$.
Besides the extra kinematical variable the analysis of the four 
fermion amplitude is more complicated  by the presence of different  
``color'' and Lorentz structures.
The Born amplitude, for example,  can be expanded in the 
basis of color/chiral amplitudes
\be
A_{B}={ig^2\over s}A^\lm
={ig^2\over s}T_F\left(-{1\over N}
\left(A_{LL}^d+A_{LR}^d\right)+A_{LL}^c+A_{LR}^c
+(L\leftrightarrow R)\right) \;,
\ee
where
\bea
A^\lm~~&=&
\bar\psi_2(p_2)t^a\gm_\mu\psi_1(p_1)
\bar\psi_4(p_4)t^a\gm_\mu\psi_3(p_3)\, , \nn \\
A_{LL}^d&=&
\bar{\psi_2}_L^i\gm_\mu{\psi_1}_L^i
\bar{\psi_4}_L^j\gm_\mu{\psi_3}_L^j \, , \\
A_{LR}^c&=&
\bar{\psi_2}_L^j\gm_\mu{\psi_1}_L^i
\bar{\psi_4}_R^i\gm_\mu{\psi_3}_R^j
\nn
\eea
and so on. 
Here $t^a$ is the $SU(N)$ generator, 
$p_1$, $p_3$ are incoming and $p_2$, $p_4$ outgoing momenta
so that $t=(p_1-p_4)^2$ and $u=(p_1+p_3)^2=-(s+t)$. 
For the moment we consider a parity conserving theory. Hence
only two chiral amplitudes are independent, for example, $LL$ and 
$LR$\footnote{A translation to the SM with the different structures of the amplitudes for
different chiralities will be considered in Section~4.}.
Similarly only two color amplitudes are independent, 
for example, $\lm$ and $d$.

Let us first compute the one loop corrections in the regime ({\em i}).
The total   contribution of  the vertex type diagrams is
\be
{\al\over \pi}
\left(C_F\left(-V_0+2V_1\right)+{C_A\over 2}V_0+\ldots\right)A_{B}\, ,
\ee
where the ellipsis stands for the 
contribution without infrared logarithms.
In the  vertex correction involving the gauge boson
selfcoupling, a contribution of the form~(\ref{v1c})   
appears with $m$ replaced by $M$ and we have used the fact that the 
pole (logarithmic) term of $V^c_1$ is the same both in eqs.~(\ref{v1c}) 
and~(\ref{v1i}).

The direct (uncrossed) box  gives
\bea
&&{ig^2\over s}{\al\over 4\pi}\bigg[\left(5B_0+3B_1+4B_2-3B_3-6B_4+2B_5\right)
\nn \\
&&\times 
\left\{\left(C_F-{T_F\over N}\right)
\bar\psi_2t^a\gm_\mu\psi_1\bar\psi_4t^a\gm_\mu\psi_3+
C_F{T_F\over N}\bar\psi_2\gm_\mu\psi_1\bar\psi_4\gm_\mu\psi_3\right\}
\nn \\
&&-\left(3B_0+B_1-B_3-2B_4+2B_5\right)
\\
&&\times \left.
\left\{\left(C_F-{T_F\over N}\right)
\bar\psi_2t^a\gm_\mu\gm_5\psi_1\bar\psi_4t^a\gm_\mu\gm_5\psi_3
+C_F{T_F\over N}\bar\psi_2\gm_\mu\gm_5\psi_1\bar\psi_4\gm_\mu\gm_5\psi_3\right\}
\right] \, ,
\nn
\eea
where $T_F=1/2$ is the index of the fundamental representation, 
\bea
B_0 = -{i}s\,J(1,1,1,1,1)\; ,~
&  B_1 = st\,J(1,2,1,1,1)\; ,
&  B_2 = st\,J(1,1,2,1,1)\, , \nn \\
B_3 = st\,J(1,3,1,1,2)\; ,~~~ 
& B_4 = st\,J(2,1,2,1,2)\; ,
&B_5 = s^2\,J(1,1,2,2,2)\; , 
\eea
and the functions $J(a_1,a_2,a_3,a_4,n)$ are proportional to scalar box
integrals with shifted dimensions:
\bea
J(a_1,a_2,a_3,a_4,n) &=& i^{-a_1-a_2-a_3-a_4-1+d/2+n}\prod_i (a_i-1)!
\nn \\ && \hspace*{-40mm} \times
\int \frac{\dd^{d+n} k }{(k^2)^{a_1}
(k^2-2p_1k-m^2)^{a_2} (k^2-2p_2k-m^2)^{a_3}(k^2-2(p_1-p_4)k+t)^{a_4}}\,.
\label{BasInt}
\eea
We apply the generalized strategy of regions to 
obtain a table of asymptotic expansions 
of the basic integrals in the leading power.
Note that, besides the hard contribution, there are two groups
of the 1c, 2c and us-contributions corresponding to two choices
of the loop momentum when it is considered to be the momentum
flowing through one of the two gauge boson lines of the given box.
Keeping terms with the leading and subleading logarithms we have
\be
B_0=B_5=B_1-B_3-2B_4=0\; ,
\ee
\bea
B^h_2(s,t)&=&-{1\over \ep^2}+{1\over \ep}\ln\left({-t}\right) 
+{1\over 2}\ln^2\left({-s}\right)-
\ln\left({-s}\right)
\ln\left({-t}\right)\; ,
\nn \\
B^c_2(s,t)&=&{1\over \ep^2}-{1\over \ep}\ln\left({-t}\right)
-\ln\left({M^2}\right)
\ln\left({-t}\right)-{1\over 2}\ln^2\left({M^2}\right)\; ,
\label{Breg}
\\
B_2(s,t)&=&{1\over 2}\ln^2\left({-s\over M^2}\right)+
\ln\left({-s\over M^2}\right)\ln\left({t\over s}\right)\;,
\nn
\eea
and the direct box contribution reads
\be
-{ig^2\over s}{\al\over \pi}B_2(s,t)
\left(\left(C_F-{T_F\over N}\right)
A^\lm
+C_F{T_F\over N}A^d\right)\; .
\ee
The crossed box  contribution can be obtained in the same way:
\be
{ig^2\over s}{\al\over \pi}B_2(s,u)
\left(\left(C_F-{T_F\over N}-{C_A\over 2}\right)
A^\lm
+C_F{T_F\over N}A^d\right) \; ,
\ee
where  $C_A=N$ is the  quadratic Casimir operator of
the adjoint  representation.
The rest of the one loop  logarithmic contributions
from the vertex corrections and the self-energy insertions 
are of the renormalization group nature.
In addition to the vertex and external fermion self-energy  contributions 
considered in the previous section the renormalization group logarithms set 
the scale of $g$ in the Born amplitude to be $Q$.

The total one loop correction  in the logarithmic
approximation reads
\[
{ig^2(Q^2)\over s}{\al\over 2\pi}
\bigg[
\left\{-C_F\ln^2\left({-s\over M^2}\right)+
\left(3C_F-C_A\ln\left({u\over s}\right)+
2\left(C_F-{T_F\over N}\right)\ln\left({u\over t}\right)
\right)
\ln\left({-s\over M^2}\right)\right\}
A^\lm 
\]
\be
\left.
+\left\{2{C_FT_F\over N}\ln\left({u\over t}\right)
\ln\left({-s\over M^2}\right)\right\}A^d
\right]\, .
\label{4fer}
\ee 
Note that the next-to-leading 
logarithms do not depend on chirality and are the same both for
the $LL$ and  $LR$ amplitudes.

Now  the collinear  logarithms can be separated from the 
total one-loop correction.
For each fermion-antifermion pair, they 
form the  exponential factor 
found in the previous section (eq.~(\ref{fi})). 
This factor in addition incorporates the   
renormalization group logarithms which are not absorbed by 
changing the normalization scale of the gauge coupling. 
The rest of the single logarithms in eq.~(\ref{4fer})
is of the soft nature.
Let us denote by   $\tilde A$   the amplitude with
the collinear logarithms  factored out.  It can be  represented as a vector
in the basis $A^\lm$, $A^d$ and satisfies the following evolution equation
\cite{Sen2,Bot} 
\be
{\partial \over \partial \ln{Q^2}}\tilde A=
{\bf \chi}(\al(Q^2))\tilde A \; ,
\label{evol3}
\ee
where  ${\bf \chi}$ is the matrix of the ``soft''
anomalous dimensions. 
 From eq.~(\ref{evol3}) we find the elements of this matrix to be, 
in units of $\al/ (4\pi)$, 
\bea
\chi_{\lm \lm} &=&
-2C_A\ln\left({u\over s}\right)+
4\left(C_F-{T_F\over N}\right)\ln\left({u\over t}\right)\; , \nn \\
\chi_{\lm d} &=&4{C_FT_F\over N}\ln\left({u\over t}\right)\; , \nn \\
\chi_{d \lm} &=& 4\ln\left({u\over t}\right)\; , \label{mat} \\
\chi_{d d} &=& 0\; .  \nn
\eea
The solution of  eq.~(\ref{evol3}) reads
\be
\tilde A= 
A^0_{1}(\al(M^2))\exp{\left[\int_{M^2}^{Q^2}{\dd x\over x}\chi_1(\al(x))\right]}+
A^0_{2}(\al(M^2))\exp{\left[\int_{M^2}^{Q^2}{\dd x\over
x}\chi_2(\al(x))\right]} \;, \ee
where $\chi_i$ are eigenvalues of the ${\bf \chi}$ matrix and 
$A^0_i$ are $Q$-independent vectors. 
Note that in higher
orders the matrices $\bf \chi$ for different values of $Q$
do not commute and the solution is given by the path-ordered exponent \cite{Sen2}.

Equations~(\ref{Breg}) imply that only the hard parts contribute
to eq.~(\ref{evol3}). This fixes the scale of $\al$ in this
equation  to be $Q$. By this reason the matrix~(\ref{mat}) 
is the same in the regime  ({\em ii}). Hence in the next-to-leading
logarithmic approximation the difference between the corrections
to the four-quark amplitude in the regimes
({\em i}) and ({\em ii}) is the one  between the factors~(\ref{fi}) and (\ref{fii}).

In the Abelian case, there are no different color
amplitudes and there is only one anomalous dimension
\be
\chi=4\ln\left({u\over t}\right)\, .
\ee

\section{Sudakov logarithms in electroweak processes}
We are interested in the process 
$f'\bar f'\rightarrow f\bar f$.
In the Born approximation, its amplitude is of   
the following form 
\be
A_{B}={ig^2\over s}\sum_{I,~J=L,~R}\left(T^3_{f'}T^3_{f}+
t^2_W{Y_{f'}Y_{f}\over 4}\right)A^{f'f}_{IJ}\; ,
\ee
where 
\be
A^{f'f}_{IJ}=\bar f_I'\gamma_\mu f_I' 
\bar f_J\gamma_\mu f_J \; ,
\ee
$t_W=\tan{\theta_W}$ with $\theta_W$ being the Weinberg angle
and  $T_f$ $(Y_f)$ is  the isospin (hypercharge) 
of the fermion which depends on the fermion chirality.

To analyze the  electroweak correction to the above process 
we use the approximation with
the $W$ and $Z$ bosons of the same mass  $M$  
and massless quarks and leptons. 
A fictitious photon mass $\lm$ has to be  introduced 
to regularize the infrared divergences.
Let us consider first the equal mass case $\lm=M$, where
we can work in terms of  the fields of 
unbroken phase and  the result of Sects.~2 and~3 for the regime ({\em i})
can be directly applied by projecting on a relevant initial/final
state.  For each fermion-antifermion  pair 
the  factor~(\ref{fi}) takes the form
\be
\exp{\left[ 
-\left(T_f(T_f+1)+t_W^2{Y_f^2\over 4}\right)\left(L(s)-
3l(s)\right)  \right]} \; ,
\label{ewfac}
\ee
where
\[
L(s)={g^2\over 16\pi^2}\ln^2\left({-s\over M^2}\right),
\]
\be
l(s)={g^2\over 16\pi^2}\ln\left({-s\over M^2}\right)\;,
\ee
and we  neglect the running
of the coupling constant in the integral in eq.~(\ref{fi}) but fix
the scale of the coupling $g$ and $t_W g$ in  
the double logarithmic contribution 
to be $Q$.
The  soft anomalous dimension for $I$ and/or $J=R$ 
is Abelian and, in units of $g^2 / (16\pi^2)$, reads 
\be
\chi=t_W^2Y_{f'}Y_f\ln\left({u\over t}\right) \;.
\label{ewab}
\ee
The  matrix of the soft anomalous dimensions for $I=J=L$ 
is a sum of the Abelian and non-Abelian parts
\bea
\chi_{\lm \lm} &=&
-4\ln\left({u\over s}\right)+
\left(t_W^2Y_{f'}Y_f+2\right)\ln\left({u\over t}\right)\; , \nn \\
\chi_{\lm d} &=&{3\over 4}\ln\left({u\over t}\right)\; , \nn\\
\chi_{d \lm} &=& 4\ln\left({u\over t}\right)\; , \\
\chi_{d d} &=& t_W^2Y_{f'}Y_f\ln\left({u\over t}\right)\; . \nn
\label{ewnab}
\eea
The photon is however massless, and 
the corresponding  infrared divergent contributions  should be
accompanied by  the real soft photon radiation
integrated to some resolution energy  $\omega_{res}$
to get an infrared safe cross 
section independent on an auxiliary photon mass. 
At the same time the massive gauge bosons are supposed
to be  detected as separate particles. 
In practice, the  resolution energy 
is much less than the $W$ ($Z$) boson mass so the   soft photon 
emission is of the QED nature. This cancels the  infrared singularities
of the QED virtual correction. We should therefore separate the QED virtual
correction from the complete result computed with the photon 
of some mass $\lm$ and then  evaluate the QED virtual corrections 
together with the real soft photon radiation effects with vanishing $\lm$. 
It is convenient to subtract the QED contribution
computed with the photon of the  mass $M$  from the obtained result
for the virtual corrections and then take the limit $\lm \to 0$ for the
sum of QED virtual and real photon contributions to the total amplitude.
In the language of the 
approach of ref.~\cite{Fad}, this prescription means that we use the  
auxiliary photon  mass $\lm$  as a variable of the evolution 
equation below the scale $M$ and the  subtraction fixes a relevant 
initial condition for this  differential equation.   
This leads to a modification of the  factor~(\ref{ewfac}) 
and the soft anomalous dimensions~(\ref{ewab}, \ref{ewnab}). 

The  common  factor for   each fermion-antifermion  pair becomes  
\be
\exp{\left[ 
-\left(T_f(T_f+1)+t_W^2{Y_f^2\over 4}-s_W^2Q_f^2\right)\left(L(s)-
3l(s)\right)  \right]} \; ,
\label{col}
\ee
where $s_W=\sin{\theta_W}$.
Then we have
\be
\chi=\left(t_W^2Y_{f'}Y_f-4s_W^2Q_{f'}Q_f\right)\ln\left({u\over t}\right)\;,
\ee
and the  matrix of the soft anomalous dimension for $I=J=L$ is
\bea
\chi_{\lm \lm} &=&
-4\ln\left({u\over s}\right)+
\left(t_W^2Y_{f'}Y_f-4s_W^2Q_{f'}Q_f+2\right)\ln\left({u\over t}\right)\; , 
\nn \\
\chi_{\lm d} &=&{3\over 4}\ln\left({u\over t}\right)\; , \nn \\
\chi_{d \lm} &=& 4\ln\left({u\over t}\right)\; , \\
\chi_{d d} &=& \left(t_W^2Y_{f'}Y_f-4s_W^2Q_{f'}Q_f\right)
\ln\left({u\over t}\right)\; . \nn
\eea
Now we can estimate the dominant one- and  two-loop logarithmic corrections.
The renormalization group logarithms which are not included into 
eq.~(\ref{col})  can be trivially taken into account by choosing 
the  relevant scale of the coupling constants in the Born amplitude. 
At the same time, the remaining logarithmic corrections 
are of the main interest because they are 
supposed to dominate  the (still unknown) total  two-loop
electroweak corrections.  
The one-loop leading and subleading  logarithms 
can be directly obtained 
from eq.~(\ref{4fer}). The corresponding corrections
to the chiral amplitudes  read 
\bea
&&\left(\left[T_f(T_f+1)+t_W^2{Y_f^2\over 4}-s_W^2Q_f^2+
(f\leftrightarrow f')\right]\left[T^3_{f'}T^3_f+
t^2_W{Y_{f'}Y_f\over 4}\right]\left(-L(s)+3l(s)\right)
\right. \nn
\\
&&+\left\{\left[-4\ln\left({u\over s}\right)
+2\ln\left({u\over t}\right)+
\ln\left({u\over t}\right)t_W^2Y_{f'}Y_f
\right]T^3_{f'}T^3_f
+{3\over 4}\ln\left({u\over t}\right)\delta_{IL}\delta_{JL}
\right.
\\
&&\left.\left.
+\ln\left({u\over t}\right)\left[t_W^2Y_{f'}Y_f-4s_W^2Q_{f'}Q_f
\right]\left[T^3_{f'}T^3_f+
t^2_W{Y_{f'}Y_f\over 4}\right]\right\}l(s)\right)
A^{f'f}_{IJ} \nn \; ,
\eea
where $\delta_{IL}=1$ for $I=L$ and zero otherwise.
The two-loop  leading (infrared)
logarithms  are determined by the
second order term  of the 
expansion of the double  (soft$\times$collinear) logarithmic 
part of the collinear  factors~(\ref{col}).
The corresponding corrections
to the chiral amplitudes  are 
\be
{1\over 2}\left(T_f(T_f+1)+t_W^2{Y_f^2\over 4}-s_W^2Q_f^2+
(f\leftrightarrow f')\right)^2
L^2(s)
A^{f'f}_{IJ}\; .
\ee
The two-loop next-to-leading   logarithms
are generated by the  interference between the first order terms
of the expansion of the
double (soft$\times$collinear) and single (soft$+$collinear$+
$re\-nor\-ma\-li\-zation group) logarithmic  exponents. 
The corresponding corrections
to the chiral amplitudes  are of the following form 
\bea
&&-\left[T_f(T_f+1)+t_W^2{Y_f^2\over 4}-s_W^2Q_f^2+
(f\leftrightarrow f')\right] \nn \\
&&\times\left\{3\left[T_f(T_f+1)+t_W^2{Y_f^2\over 4}-s_W^2Q_f^2+
(f\leftrightarrow f')\right]\left[T^3_{f'}T^3_f+
t^2_W{Y_{f'}Y_f\over 4}\right]
\right. \nn
\\
&&+\left[-4\ln\left({u\over s}\right)
+2\ln\left({u\over t}\right)+
\ln\left({u\over t}\right)t_W^2Y_{f'}Y_f
\right]T^3_{f'}T^3_f
+{3\over 4}\ln\left({u\over t}\right)\delta_{IL}\delta_{JL}
 \\
&&\left.
+\ln\left({u\over t}\right)\left[t_W^2Y_{f'}Y_f-4s_W^2Q_{f'}Q_f
\right]\left[T^3_{f'}T^3_f+
t^2_W{Y_{f'Y_f}\over 4}\right]\right\}L(s)l(s)
A^{f'f}_{IJ}\nn \; .
\eea
With the expression for the  chiral amplitudes
at hand, we can compute the leading and subleading logarithmic corrections
to the basic observables for $e^+e^-\rightarrow f \bar f$
using standard formulae.
Though the above approximation is not formally valid for small angles 
$\theta <M/\sqrt{s}$   
we can integrate the  differential cross section 
over all angles to get a result with the logarithmic accuracy. 
Let us, for example, consider  the total cross sections
of the quark-antiquark/$\mu^+\mu^-$ production in 
the $e^+e^-$ annihilation.
In the two loop approximation, the  leading and next-to-leading
Sudakov  corrections to the cross sections read
\bea
\sigma/\sigma_B(e^+e^-\rightarrow Q\bar Q)\hspace{3mm}
&=&1+\hspace{2mm}5.30\,l(s)-1.66\,L(s)-12.84\,l(s)L(s)+1.92\,L^2(s) \; ,
\nn \\
\sigma/\sigma_B(e^+e^-\rightarrow q\bar q)\hspace{6mm}
&=&1+20.54\,l(s)-2.17\,L(s)-53.72\,l(s)L(s)+2.79\,L^2(s)\; ,
\label{sig}
\\
\sigma/\sigma_B(e^+e^-\rightarrow \mu^+\mu^-)
&=&1+10.09\,l(s)-1.39\,L(s)-21.66\,l(s)L(s)+1.41\,L^2(s)\; ,
\nn
\eea
where $Q=u,c,t$,  $q=d,s,b$, and we use $s_W^2=0.232$ for numerical
estimates. 
Numerically, $L(s)=0.07$  $(0.11)$ and $l(s)=0.014$  $(0.017)$
respectively for $\sqrt{s}=1$~TeV and $2$~TeV. 
Here $M=M_W$
has been chosen for the infrared cutoff and $g^2/16\pi^2=2.7\cdot 10^{-3}$
for the $SU(2)$ coupling. 
 For physical applications, the running of $g$ and $s_W$, the  
$W$ and  $Z$ boson mass difference
and the top quark mass effects are important in the one loop 
approximation. Fortunately  the first order corrections are known exactly 
beyond the  logarithmic approximation \cite{Hol}. 

Clearly, for energies at 1 and 2 TeV
the two loop corrections are huge and amount up respectively
to $5\%$ and $7\%$. 
There is a cancellation between the leading and 
subleading logarithms and for the above energy interval the
subleading contribution even exceeds the  leading one.
The higher order leading and next-to-leading
corrections however do not exceed $1\%$ level.
They can be in principle resummed using the formulae given above.

 For completeness we give a numerical estimate
of corrections to the  cross section asymmetries. 
In the case of  the  forward-backward  asymmetry (the difference of 
the cross section averaged over forward and backward semispheres 
with respect to the electron beam direction
divided by the total cross section) we get
\bea
A_{FB}/A_{FB}^B(e^+e^-\rightarrow Q\bar Q)\hspace{3mm}
&=&1-1.23\,l(s)-0.09\,L(s)+0.11\,l(s)L(s)+0.12\,L^2(s)\; ,
\nn \\
A_{FB}/A_{FB}^B(e^+e^-\rightarrow q\bar q)\hspace{6mm}
&=&1+7.16\,l(s)-0.14\,L(s)-1.54\,l(s)L(s)+0.02\,L^2(s)\; ,
\nn \\
A_{FB}/A_{FB}^B(e^+e^-\rightarrow \mu^+\mu^-)
&=&1+5.48\,l(s)-0.04\,L(s)-6.39\,l(s)L(s)+0.27\,L^2(s)\; .
\nn 
\eea
 For the left-right asymmetry (the difference of the cross sections 
of the left and right particles production divided by the 
total cross section) we obtain
\bea
A_{LR}/A_{LR}^B(e^+e^-\rightarrow Q\bar Q)\hspace{3mm}
&=&1+\hspace{2mm}10.17\,l(s)-\hspace{2mm}
2.77\,L(s)+ \hspace{2mm}
6.38\,l(s)L(s)-0.91\,L^2(s)\; ,
\nn \\
A_{LR}/A_{LR}^B(e^+e^-\rightarrow q\bar q)\hspace{6mm}
&=&1+\hspace{2mm}11.66\,l(s)-\hspace{2mm}
1.08\,L(s)+15.27\,l(s)L(s)-0.77\,L^2(s)\; ,
\nn \\
A_{LR}/A_{LR}^B(e^+e^-\rightarrow \mu^+\mu^-)
&=&1+118.07\,l(s)-13.74\,L(s)-\hspace{2mm}
1.13\,l(s)L(s)-0.78\,L^2(s)\; .
\nn 
\eea
Let us compare our results with results of previous analyses.  
Our result for the one loop  double logarithmic
contribution is in agreement with \cite{CiaCom}. However
the result for the   one loop single  infrared logarithmic
contribution differs from \cite{Bec}. The reason is that, in \cite{Bec},
only the diagrams with heavy virtual bosons have been taken into
account.  There is an infrared safe contribution of the diagram with 
the virtual massless photon where the heavy boson
mass serves as an infrared regulator that should be taken into
account to get a complete (exponential) result. 
In one-loop approximation,
this contribution comes from the box diagrams with
the photon and $Z$ boson running inside the loop.
One of the two collinear regions of these diagrams 
(see Section 3) gives an infrared safe contribution
that should be taken into account.   
The contribution from this diagram combined with real radiation has been
obtained in analytical form in \cite{KS}.
Note that   neglecting  contributions of this type leads also to
the breakdown of the exponentiation of the double logarithms starting
from the two-loop approximation \cite{KuhPen}.

Our result for the two-loop  double logarithmic
contribution is in agreement with \cite{Fad}.
On the other hand, the coefficients in front of the two-loop
leading logarithms in eq.~(\ref{sig}) with a few percent
accuracy coincide  with the result of \cite{KuhPen} where
the photon contributions
were not considered. This is related to the fact 
that  the virtual photon contribution not included to the result
of  \cite{KuhPen} is suppressed by a small factor $s_W^2$. 

Note that our analysis implies the resolution energy  
for the real photon
emission to be smaller than the heavy boson mass. 
The above result for the cross section should be multiplied
by the standard factor which takes into account the soft photon emission 
and the {\em pure} QED virtual corrections. This factor 
depends logarithmically on  $s$, $\omega_{res}$ and the initial/final
fermion masses but not on $M_{Z,W}$.
If the resolution energy exceeds $M_{Z,W}$ the analysis is more
complicated due to  the fact that the radiation of real photons
is not of Poisson type because of its non-Abelian $SU(2)_L$ component.
A complete analysis of this problem  in the double 
logarithmic approximation is given in ref.~\cite{Fad}.  
However, as we have pointed out the effects
of the non-Abelian component of the photon are numerically 
rather small.

\section{Conclusion}
In the present paper,  we have analyzed 
the asymptotic behavior of  the Abelian form factor 
and four fermion amplitude in the $SU(N)$ gauge theory 
in two standard variants of  the Sudakov limit:
with on-shell massless fermions and massive gauge 
bosons and with  off-shell massless fermions and massless gauge 
bosons.   The generalized strategy of regions and 
dimensional regularization
were used to obtain  the asymptotic expansions of one-loop diagrams
that determine  the  structure of  the evolution equations
for the amplitudes in the Sudakov limit  up to the  next-to-leading 
logarithmic approximation. 
By integrating these equations the next-to-leading 
logarithms were summed up.
The  method can be directly extended to   
the next-to-next-to-leading logarithms.
To do this  one can apply 
expansions of two-loop Feynman integrals
within the strategy of regions (in the case of the form factor
see examples of expansions of master scalar integrals in 
\cite{SR,Smr}) and insert two-loop information 
into the evolution equations.

We have applied our  results  
to the analysis of the   electroweak  corrections  to the 
process of fermion-antifermion pair production
in the $e^+e^-$ annihilation. The two-loop leading and 
next-to-leading logarithmic corrections to the chiral amplitudes
which are supposed to saturate the total
two-loop electroweak corrections in TeV region
have been obtained. 
The corresponding corrections to the  
total cross sections and asymmetries of the quark-antiquark 
and $\mu^+\mu^-$ 
production in the $e^+e^-$ annihilation have been found  
to be of a few percent magnitude at the energy of 1-2 TeV.
The next-to-leading  infrared logarithms are comparable
and even exceed the  leading  ones at this scale.

\vspace{4mm}

{\bf Acknowledgments}\\[3mm]
The work by V.S. was supported by the Volkswagen Foundation, contract
No.~I/73611. The work by J.K. and A.P. was supported by the Volkswagen
Foundation, by BMBF under grant BMBF-057KA92P and by DFG-Forschergruppe
``Quantenfeldtheorie, Computeralgebra und Monte-Carlo-Simulationen''
(DFG Contract KU502/6--1). The work by  A.P. was supported by
the Bundesministerium f\"ur Bildung und Forschung
under Contract No.\ 05~HT9GUA~3, and by the European Commission through the
Research Training Network {\it Quantum Chromodynamics and the Deep Structure
of Elementary Particles} under Contract No.\ ERBFMRXCT980194.
We are greatful to M.~Beneke for careful reading of the manuscript.
V.S. is thankful to the institute of Theoretical 
Particle Physics of the University of Karlsruhe for kind
hospitality when this work was completed.


\begin{thebibliography}{99}

\bibitem{Sud}
V.V. Sudakov, {\em Zh. Eksp. Teor. Fiz.} 30 (1956) 87.

\bibitem{Jac} R. Jackiw, {\em Ann. Phys.} 48 (1968) 292; 51 (1969) 575.

\bibitem{CorTik} J.M. Cornwall and  G. Tiktopoulos, {\em Phys. Rev. Lett.}
35 (1975) 338;
{\em Phys. Rev.} D13 (1976) 3370.

\bibitem{FreTay} J. Frenkel and J.C. Taylor, {\em  Nucl. Phys.}
B116 (1976) 185.

\bibitem{Smi}  V. Smilga {\em Nucl.Phys. } B161 (1979) 449.

\bibitem{Col}
J.C. Collins, {\em Phys.Rev. } D22 (1980) 1478;
in {\em Perturbative QCD}, ed. A.H. Mueller, 1989, p.~573.

\bibitem{Sen1} A. Sen, {\em Phys. Rev. } D24 (1981) 3281.

\bibitem{Sen2} A. Sen, {\em Phys. Rev. } D28 (1983) 860.

\bibitem{Kor1} G. Korchemsky, {\em Phys. Lett. } B217 (1989) 330.

\bibitem{Kor2} G. Korchemsky, {\em Phys. Lett. } B220 (1989) 629.

\bibitem{Mag} L.~Magnea and G.~Sterman, {\em Phys. Rev. } D42 (1990) 4222.

\bibitem{BS}
M. Beneke and V.A.~Smirnov, {\em Nucl. Phys.} B522 (1998) 321.

\bibitem{SR}
V.A.~Smirnov and E.R. Rakhmetov, {\em Teor. Mat. Fiz.} 120 (1999) 64;
V.A.~Smirnov, {\em Phys. Lett. } B465 (1999) 226.

\bibitem{dimreg}
G.~'t Hooft and M.~Veltman, {\em Nucl.~Phys.} B44 (1972) 189;
C.G.~Bollini and J.J.~Giambiagi, {\em Nuovo Cim.} 12B (1972) 20.

\bibitem{KDB} M. Kuroda, G. Moultaka and D.~Schildknecht, {\em Nucl. Phys.}
              {B350} (1991) 25;\\
              G.~Degrassi and A.~Sirlin, {\em Phys. Rev.} {D46} (1992) 3104;\\
              M.~Beccaria {\it et al.}, {\em Phys. Rev.} {D58} (1998) 093014. 

\bibitem{CiaCom} P.~Ciafaloni and  D.~Comelli,  {\em Phys. Lett.} {B446} 
                 (1999) 278.

\bibitem{Bec}    M.~Beccaria {\em et al.}, Preprint {PM/99--26}, 
                 hep-ph/9906319. 

\bibitem{KuhPen} J.H. K\"uhn and A.A. Penin, Preprint TTP/99--28,
                 hep-ph/9906545.

\bibitem{Fad} V.S.~Fadin, L.N.~Lipatov, A.D.~Martin and M.~Melles,
              Preprint  PSI PR--99--24, hep-ph/9910338.

\bibitem{Ste} G.~Sterman {\em Phys. Rev. } D17 (1978) 2773;\\
              S.Libby and G.~Sterman {\em Phys. Rev. } D18 (1978) 3252;\\
              A.H. Mueller, {\em Phys. Rep.} 73 (1981) 35.

\bibitem{FreMeu} J. Frenkel and R. Meuldermans,  {\em Phys. Lett.} 
                 {B65} (1976) 64.

\bibitem{Fre}  J.~Frenkel,  {\em Phys. Lett.} {B65} (1976) 383.

\bibitem{Ama}   D. Amati, R. Petronzio and G.~Veneziano,  
                {\em Nucl. Phys.} B146 (1978) 29.


\bibitem{Bot} J.~Botts and G.~Sterman, {\em Nucl. Phys. } B325 (1989) 62;\\
              N.~Kidonakis, G~Odereda and G.~Sterman,
              {\em Nucl. Phys. } B531 (1998) 365.

\bibitem{Hol} W.~Beenakker, W.~Hollik and Van der Mark, 
               {\em Nucl. Phys.} {B365} (1991) 24.

\bibitem{KS} J.H. K\"uhn and R.G.~Stuart,  
{\em Phys. Lett. } B200 (1987) 360.

\bibitem{Smr}
V.A.~Smirnov, {\em Phys. Lett. } B404 (1997) 101.


\end{thebibliography}
\end{document}